\documentclass[a4paper]{revtex4}
\usepackage{amssymb}
\usepackage{amsmath}
\usepackage{graphicx}

\usepackage[english]{babel}
\usepackage[T1]{fontenc}
\usepackage[utf8]{inputenc}

\begin{document}

\newcommand{\Ec}{\mathcal{E}}
\newcommand{\ec}{{\varepsilon}}
\newcommand{\ep}{{\epsilon}}

\newcommand{\Kf}{\ifmmode K^2_{1/3} \else ERROR \fi}
\newcommand{\Ks}{\ifmmode K^2_{2/3} \else ERROR \fi}
\newcommand{\Kfs}{\ifmmode K_{1/3}K_{2/3} \else ERROR \fi}

\newcommand{\Kone}{\ifmmode K_{1/3} \else ERROR \fi}
\newcommand{\Ktwo}{\ifmmode K_{2/3} \else ERROR \fi}

\title{Depolarization of synchrotron radiation of a relativistic electron beam}
\author{O.~Novak}
\author{M.~Diachenko}
\author{\fbox{R.~Kholodov}}
\affiliation{Institute of applied physics, National academy of sciences of Ukraine, Petropavlivska street, 58, 40000 Sumy, Ukraine}
\date{2025}

\begin{abstract}
We present a theoretical study on the radiative self-polarization of a high-energy electron beam propagating perpendicular to a strong magnetic field.
Recently, a similar setup has been proposed as a source of polarized electron and photon beams.
We focus on the dependence of electron and radiation polarization on the dimensionless parameter $\ec$, which is proportional to the product of electron energy and magnetic field strength.
The numerical solution of the balance equation shows that the resulting electron beam polarization increases rapidly as a function of $\ec$ for $\ec \ll 1$ and saturates at a value of approximately $-0.8$.
If $\ec \gg 1$, the rate of self-polarization decreases significantly.
At the same time, a substantial or nearly complete depolarization of synchrotron radiation is observed, particularly for an electron beam with spins initially aligned parallel to the field.
\end{abstract}

\maketitle

\section{Introduction}
\label{sec:intro}
Radiative self-polarization of relativistic electrons, or Sokolov and Ternov effect, was theoretically predicted in Ref.~\cite{Sokolov64}, and later observed experimentally.
The effect consists in the gradual alignment of spins of electrons moving in a magnetic field.
This behavior can be explained by the possibility of a change in spin projection during synchrotron radiation, with a higher probability of transition to a spin-down state.
From its discovery to the present days, radiative selfpolarization was investigated in a wast number of works (see, for expample, \cite{Sokolov, Mane97,Meot21} and references therein).
At the same time, most studies consider the effect under conditions typical for existing accelerators and storage rings.
In practice, achievable magnetic fields are much weaker than the critical Schwinger field, $H_c \approx 4.41\cdot 10^{13}$~G, and the spectral maximum of synchrotron radiation is located at a frequency much smaller than the electron energy.
Moreover, the energy lost to radiation is compensated by the storage ring, which keeps electrons close to the specified trajectory.
If the electron energy remains constant, it is possible to introduce the asymptotic equilibrium value of the polarization degree for electrons, which equals approximately to 92\%~\cite{Sokolov64, Sokolov}.

The self-polarization time can range from minutes to even hours in practice, but it decreases rapidly with increasing field strength.
It is expected that in a strong field the time of radiative self-polarization falls within the femtosecond range.
In laboratory conditions, the strongest feasible field can be generated by lasers, which raised considerable interest in Sokolov-Ternov effect in an electromagnetic wave.
However, observing the selfpolarization in laser wave is complicated due to precession of spin in the field of electromagnetic wave.
Among the proposed solutions, considering the polarization of the beam in the radial plane, the use of a bichromatic or counterpropagating laser beams, and polarization due to target interaction were suggested, in particular~\cite{Buescher20, DelSorbo17, DelSorbo18, Seipt18, Seipt19, Li19, Shen24}.

At the same time, advances in experimental techniques open ways to a significant increase in the magnetic field strength, available in laboratory conditions.
In Ref.~\cite{Zhu23b, Zhu23a, Xue24, Zhu24}, authors propose an approach to generate magnetic field in the gigagauss range.
In this approach, dense electron bunch collides with a solid target positioned at a grazing angle to the bunch velocity.
Before the electron bunch collides with the target surface, a strong return electron current is swiftly induced in the target plate to maintain current neutralization. This results in the prompt formation of gigagauss magnetic fields, which can confine and focus the injected electron beam~\cite{Xue24}.
As particle-in-cell simulation shows, the magnetic field strength can reach values up to 4~GG for the beam parameters, feasible in the FACET II experiment at SLAC \cite{FACETII}.
The resulting magnetic field is determined by the beam energy and bunch size and charge, and spans the spatial region comparable in size to the bunch.

In strong field experiments, the self-polarization regime will differ from that in storage rings.
The role of quantum effects considerably increases in strong field, which necessitates the use of accurate expressions for the rate of synchrotron transitions.
Another important aspect is the free propagation of the electron bunch without the influence of a storage ring.
Thus, it is necessary to take into account the energy losses of electrons and the evolution of the electron energy distribution function.

In the present work we consider hypothetical experimental setup with free passage of an electron beam through a region with a strong magnetic field.
For simplicity we assume the magnetic field to be constant and uniform field directed perpendicular to the beam momentum.
At the same time, we do not apply any restrictions on the field strength and electron energy, apart from the general requirement of relativistic beam energy, and the Larmour radius being large compared to the electron path in the field.
To account for radiative energy losses and multiple synchrotron transitions, we track the evolution of the electron energy distribution for each spin component of the electron beam by numerically solving the corresponding balance equation.
The main attention is focused on the radiative self-polarization of the beam, and on the polarization of the accompanying synchrotron radiation.
In particular, we investigate the time evolution of the electron polarization degree and the Stokes parameters of synchrotron radiation, and their dependence on the dimensionless parameter proportional to the beam energy and magnetic field strength.

\section{Synchrotron rates}
Synchrotron radiation has been extensively studied in the scientific literature in the past~\cite{Sokolov, Klepikov54, Chiu69, Tsai73, Latal79, Harding87, Hofmann08, Novak08, Novak09}.
For the sake of convenience, we recite the expression of the differential rate of synchrotron radiation.
An electron in a magnetic field occupies discrete energy states, characterized by specific values of energy, the longitudinal momentum, and the spin projection along the field direction.
Electron spin state can be conveniently characterized by the electron polarization defined as $\mu = 2s_z$, with $\mu = \pm 1$.
If the electron energy significantly exceeds the spacing between energy levels, it can be considered as a continuous quantity, as well as the frequency of the emitted photon.
In this case, the significance of strong-field quantum effects can be quantified using the quantum nonlinearity parameter defined as~\cite{Fedotov23}
\begin{equation}
\label{genInvariant} 
  \ec = \frac{\left| F_{\mu\nu} p^\nu\right|}{mc \: H_c},
\end{equation}
where $H_c = m^2c^3/e\hbar$ is the critical Schwinger field, $H_c \approx 4.41\cdot 10^{13}$~G.
For a relativistic particle moving perpendicular to the magnetic field, this parameter reduces to
\begin{equation}
\label{epsilon} 
  \ec = \frac{\Ec}{mc^2}\frac{H}{H_c}.
\end{equation}
By introducing dimensionless time $\tau$ as 
\begin{equation}
\label{tau}
    \tau = \alpha \omega_H t,
\end{equation}
where $\alpha$ is the fine structure constant and $\omega_H = eH/mc$ is the cyclotron frequency, the synchrotron rates can be expressed solely in terms of the invariants $\ec$.
Thus, the differential rate of synchrotron transition from the state $(\ec, \mu)$ to the energy interval $d\ec'$ in the vicinity of the state $(\ec',\mu')$ can be written as
\begin{equation}
\label{rates}
    dw^{\mu'\mu}(\ec', \ec) = \frac{1}{16\pi\sqrt{3}} \frac{(\ec+\ec')^2}{\ec^3\ec'} G^{\mu'\mu} d\ec',
\end{equation}
\begin{equation}
\label{Gmm}
    \begin{array}{ll}
        G^{\mu\mu} &= 
        \left( \Ktwo(a)-Y \right) (1+\xi) + 
        \left( 3\Ktwo(a) + (2\rho^2-1)Y(a) - 4\mu\rho \Kone(a) \right) (1-\xi),
        \\
        G^{-\mu\mu} &= \rho^2 \left[
        \left(\Ktwo(a)-Y(a)\right)(1-\xi) + 
        \left( 3\Ktwo(a) + Y(a) + 4\mu \Kone(a) \right) (1+\xi)
        \right],
    \end{array}
\end{equation}
where
\begin{equation}
    \rho = \frac{\Omega}{\ec+\ec'},
\end{equation}
\begin{equation}
    a = \frac23 \frac{\Omega}{\ec\ec'} = \frac{4}{3\ec} \frac{\rho}{1-\rho}.
\end{equation}
In the above expressions, $\Omega = \ec - \ec'$ is the dimensionless photon frequency, $K_v(a)$ are the modified Bessel functions, and $Y(a)$ is the integral Bessel function defined as 
\begin{equation}
    Y(a) = \int\limits_a^\infty \Kone(x)dx.
\end{equation}

The synchrotron rates given by Eqs.~(\ref{rates})--(\ref{Gmm}) are integrated over the emission angle and depend on a single Stokes parameter $\xi$, which describes the photon polarization in the horizontal and vertical directions with respect to the magnetic field.
By convention, the coordinate system is chosen so that $\xi = -1$ corresponds to photon polarization perpendicular to the plane spanned by the magnetic field vector $\vec H$ and the photon wave vector $\vec k$, and  $\xi = +1$ corresponds to the photon polarization laying within the plane $(\vec H, \vec{k})$.
The Stokes parameter $\xi$ determines the probability of observing the emerging photon in the corresponding polarization state, and should be interpreted as related to the analyzer setup description.

If we are not intrested in photon polarization, rates (\ref{rates}) must be integrated over $\xi$. 
The trivial integration leads to
\begin{equation}
\label{unpolarized}
    w^{\mu'\mu}(\ec', \ec) = 2w^{\mu'\mu}(\ec', \ec, \xi)|_{\xi=0}.
\end{equation}

The total rate of synchrotron emission as a function of the dimensionless energy $\ec$ shows power-law decay for $\ec \gg 1$.
Therefore, one may expect reduction in self-polarization speed for greater values of $\ec$.

\section{Balance equation}
To investigate the evolution of electron density of spin components of the electron beam we consider the number of electrons with polarization $\mu$ and their energy energy $\ec$ within the small interval $d\ec$.
From the definition of the electron density $n_\mu(\tau, \ec)$ we can find the number of electrons in such interval as $dN_\mu(\tau, \ec)=n_\mu(\tau, \ec)d\ec$.
This number decreases due to synchrotron transitions to states with lower enrgy $\ep<\ec$ and the final electron polarization $\nu = \pm\mu$.
Given the synchrotron rates~Eq.~(\ref{rates}), the number of electrons leaving the interval $d\ec$ during time span $d\tau$ can be written as
$dN_\mu(\tau, \ec)w^{\nu\mu}(\ep, \ec)d\ep d\tau$ with $w^{\mu'\mu}(\ec', \ec)$ given by (\ref{unpolarized}).
Finally, after integrating over the final energy $\ep$ and summing over the final polarization $\nu=\pm\mu$, the total electron loss due to radiation reads
\begin{equation}
\label{eloss}
    \delta N^-_\mu(\tau,\ec) = d\ec d\tau \: n_\mu(\tau, \ec) \sum\limits_{\nu=\pm\mu} \int\limits_0^\ec  w^{\nu\mu}(\ep, \ec)d\ep.
\end{equation}
Similarly, synchrotron transitions from states $(\ep, \nu)$ with $\ep>\ec$ cause inflow of electrons to the energy interval $d\ec$.  
Following similar reasoning, the increase in electron number can be written as 
\begin{equation}
\label{einflow}
    \delta N^+_\mu(\tau, \ec) = d\ec d\tau \: 
    \sum\limits_{\nu=\pm\mu} \int\limits_\ec^\infty  n_\nu(\tau, \ep) w^{\mu\nu}(\ec, \ep)d\ep.
\end{equation}
Subtracting (\ref{einflow}) from (\ref{eloss}) and dividing by $d\ec d\tau$, one can find the following balance equation describing the evolution of the electron energy density,
\begin{equation}
\label{balance}
     \frac{\partial n_\mu(\tau,\ec)}{\partial \tau} = \sum\limits_{\nu=\pm\mu}
     \left[
        \int\limits_\ec^\infty n_\nu(\tau, \ep)    w^{\mu\nu}(\ec, \ep)d\ep   - 
        n_\mu(\tau, \ec) \int\limits_0^\ec  w^{\nu\mu}(\ep, \ec)d\ep
     \right].
\end{equation}
As any processes that change total electron number are not included in the consideration, the energy density of spin components satisfies the  normalizing condition
\begin{equation}
\label{norm}
    \int_0^\infty \left( \vphantom{\int_I^I}
        n_{_+}(t,\ec) + n_{_-}(t,\ec) \right) d\ec  = N,
\end{equation}
where $N$ is the total electron number.

The rates $w^{\mu\mu}(\ec',\ec)$ exhibit divergences as $\ec' \rightarrow \ec$ due to the presence of singular special functions $K_\nu(a)$ and $Y(a)$.
Although these divergences are integrable and do not pose essential difficulties, special attention must be paid to the choice of numerical procedure used to solve Eq.~(\ref{balance}).
To correctly handle the singularities, the tanh-sinh and exp-sinh quadratures were employed in the present work~\cite{Takahashi74}.

The equation (\ref{balance}) should be supplemented with the initial energy distribution.
In this paper we present results based on the numerical solution of the balance equation (\ref{balance}) for the beam with the normal energy distribution at $\tau=0$.
The beam energy $\ec_0$ ranges from $10^{-2}$ to $10^{5}$, and the standard deviation $\sigma$ is set to $0.1\ec_0$ in all cases.
The numerical calculations show that polarization of the beam and its radiation does not show strong dependence on the value of $\sigma$.
The cases of full alignment of electron spins parallel and antiparallel to the magnetic field, as well as an initially unpolarized beam,  were considered.

\begin{figure}
    \includegraphics[width=0.49\columnwidth]{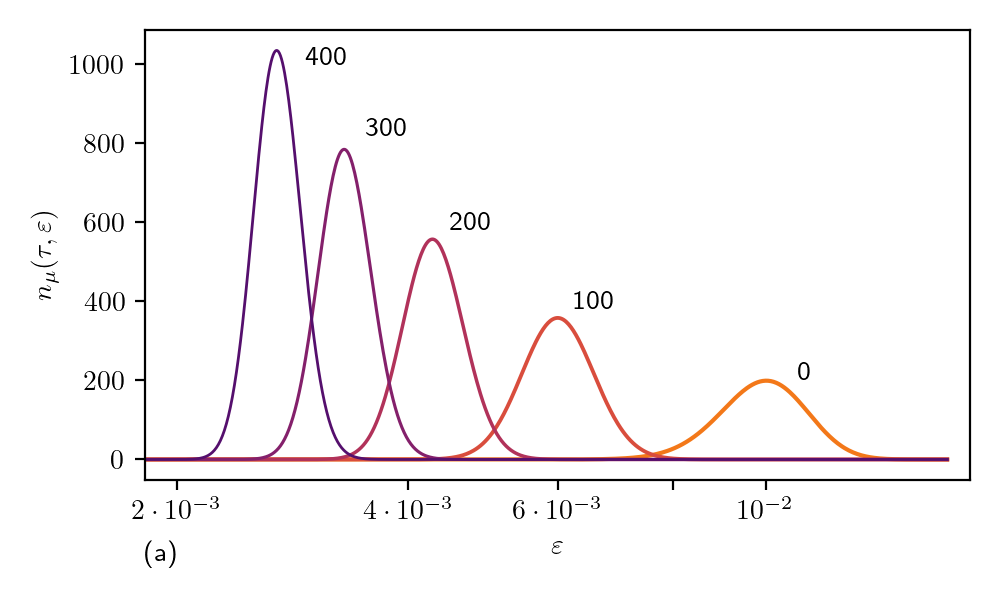} 
    \includegraphics[width=0.49\columnwidth]{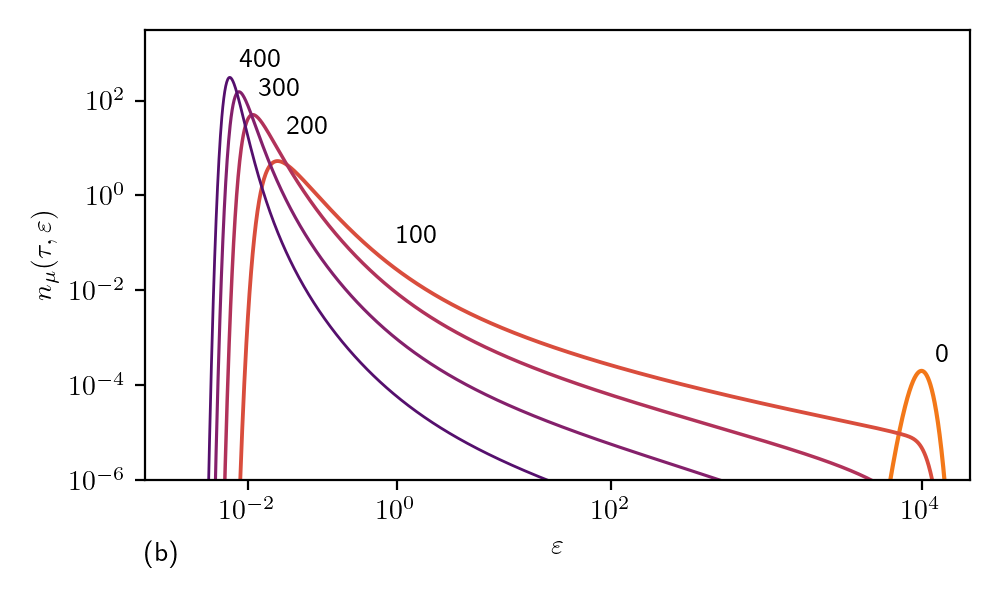}
  \caption{Evolution of the electron density of the spin component $\mu=-1$. The initial beam energy is (a) $\ec_0=10^{-2}$ and (b) $\ec_0=10^4$. Curves correspond to different moments of the dimensionless time parameter: $\tau = 0$; 100; 200; 300; 400. }
  \label{fig:nevo}
\end{figure}

The behaviour of the electron density $n_\mu(\tau, \ec)$ considerably depends on the mean beam energy, as can be seen from Fig.~\ref{fig:nevo}.
If the energy is small, $\ec_0 \ll 1$, then emission of low frequency photons dominates, and the density distribution gradually shifts to lower energy values.
On the contrary, high energy electrons with $\ec \gg 1$ emit photons with frequency $\Omega \sim \ec$, resulting in the emergence and growth of a local maximum in the low-energy range.

\section{Electron polarization degree}
Having obtained the electron density from the solution of the balance equation, we may proceed to investigation of the spin and polarization properties.
The spin alignment of an electron beam can be characterized by a polarization degree defined as
\begin{equation}
\label{Pe}
  P_{e} = \frac{N_+ - N_-}{N_+ + N_-}  ,
\end{equation}
where $N_\pm$ is the electron number with corresponding polarization $\mu = \pm 1$.
Thus, $P_e=+1$ when electron spins are aligned parallel to the magnetic field, $P_e=-1$ when they are antiparallel to the field, and $P_e=0$ for an unpolarized beam.
Taking into account conservation of the total electron number and electron density definition, $P_e$ can be expressed as
\begin{equation}
    P_e(\tau) = \frac{1}{N} \int\limits_0^\infty \left(n_+(\tau, \ec) - n_-(\tau, \ec)\right) d\ec.
\end{equation}

\begin{figure}
  \includegraphics[width=0.85\columnwidth]{{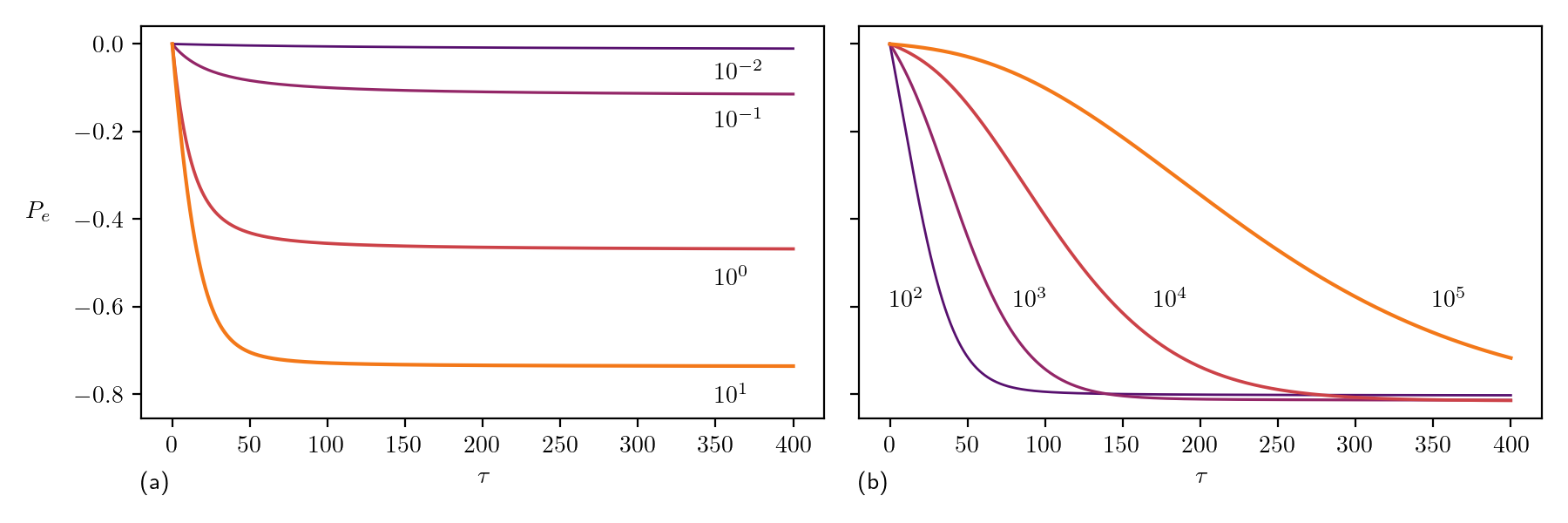}}
  \caption{Polarization degree of an initially unpolarized beam as a function of time for different values of the initial dimensionless beam energy: (a) $\ec_0=10^{-2}$, $10^{-1}$, $10^{0}$, $10^{1}$; (b) $\ec_0=10^{2}$, $10^{3}$, $10^{4}$, $10^{5}$. }
  \label{fig:PvsE0}
\end{figure}

Figure~\ref{fig:PvsE0} shows the electron polarization degree $P_e$ as a function of the dimensionless time $\tau$ in the case of an initially unpolarized beam.
Two stages of radiative self-polarization can be clearly distinguished. 
At the beginning, the absolute value of the polarization degree increases rapidly, accompanied by significant energy losses due to radiation.
During the following time interval the beam energy is low and the polarization degree changes only slightly.
Note that in the case under consideration, the beam energy is not replenished by a storage ring and decreases steadily. Therefore, the asymptotic value of the polarization degree does not necessarily exist.

When the beam energy is small, $\ec_0 \ll 1$, the resulting polarization degree at the end of simulation increases with $\ec_0$, and it reaches the maximum value of $\sim 80$\%.
In contrast, the behavior of the polarization degree differs in the range $\ec_0 \gg 1$.
The final value of the polarization degree practically does not change, while the speed of radiative selfpolarization decreases.
For instance, when $\ec_0 = 10^5$, the dimensionless time corresponding to the duration of the first stage is by an order of magnitude longer in comparison with $\ec_0 = 10$.

\section{Radiation polarization}
Before investigating the polarization of radiation, we note that, in general, the polarization of synchrotron radiation depends on the emission angle and radiation frequency.
However, relativistic particles radiate predominantly into the narrow cone along the direction of motion, with the cone opening angle about $\sim mc^2/\Ec$. 
In the present work, we assume that all radiation passes entirely through the analyzer aperture and, hence, we do not consider the angle dependence of polarization.

Let $Q$ and $\xi$ be the Stokes parameters describing the radiation polarization and the analyzer configuration respectively.
Then, $Q$ are determined by the difference in intensity measured with opposite configurations of the analyzer device corresponding to  $\xi=\pm1$,
\begin{equation}
\label{stokes}
  Q = \frac{I(\xi = +1) - I(\xi = -1)}{I(\xi = +1) + I(\xi = -1)}.
\end{equation}

The differential radiation intensity of a single electron is given by the transition rate multiplied by the radiation frequency $\Omega$.
To obtain the radiation intensity of an electron beam described by the energy distributions $n_\mu(\tau, \ec)$, one must sum the contributions of all electrons with energy satisfying $\ec \geq \Omega$ taking into account both spin components of the beam,
\begin{equation}
\label{intensity}
    I(\tau, \Omega, \xi) = \Omega \sum\limits_{\mu'\mu}\int\limits_\Omega^\infty
    n_\mu(\tau, \ep) w^{\mu'\mu}(\ep-\Omega,\ep,\xi)d\ep,
\end{equation}
where $w^{\mu'\mu}(\ec',\ec, \xi)$ are given by Eqs.~(\ref{rates})--(\ref{Gmm}).
Inserting (\ref{intensity}) into (\ref{stokes}) allows us to compute the Stokes parameter $Q$ as a function of time and frequency.
If the frequency dependence of the polarization is not of intrest, the intensity (\ref{intensity}) should be also integrated over $\Omega$.
Obviously, the other Stokes parameters vanish since the intensity (\ref{intensity}) depends on the single parameter $\xi$.

\begin{figure}
  \includegraphics[width=1.0\columnwidth]{{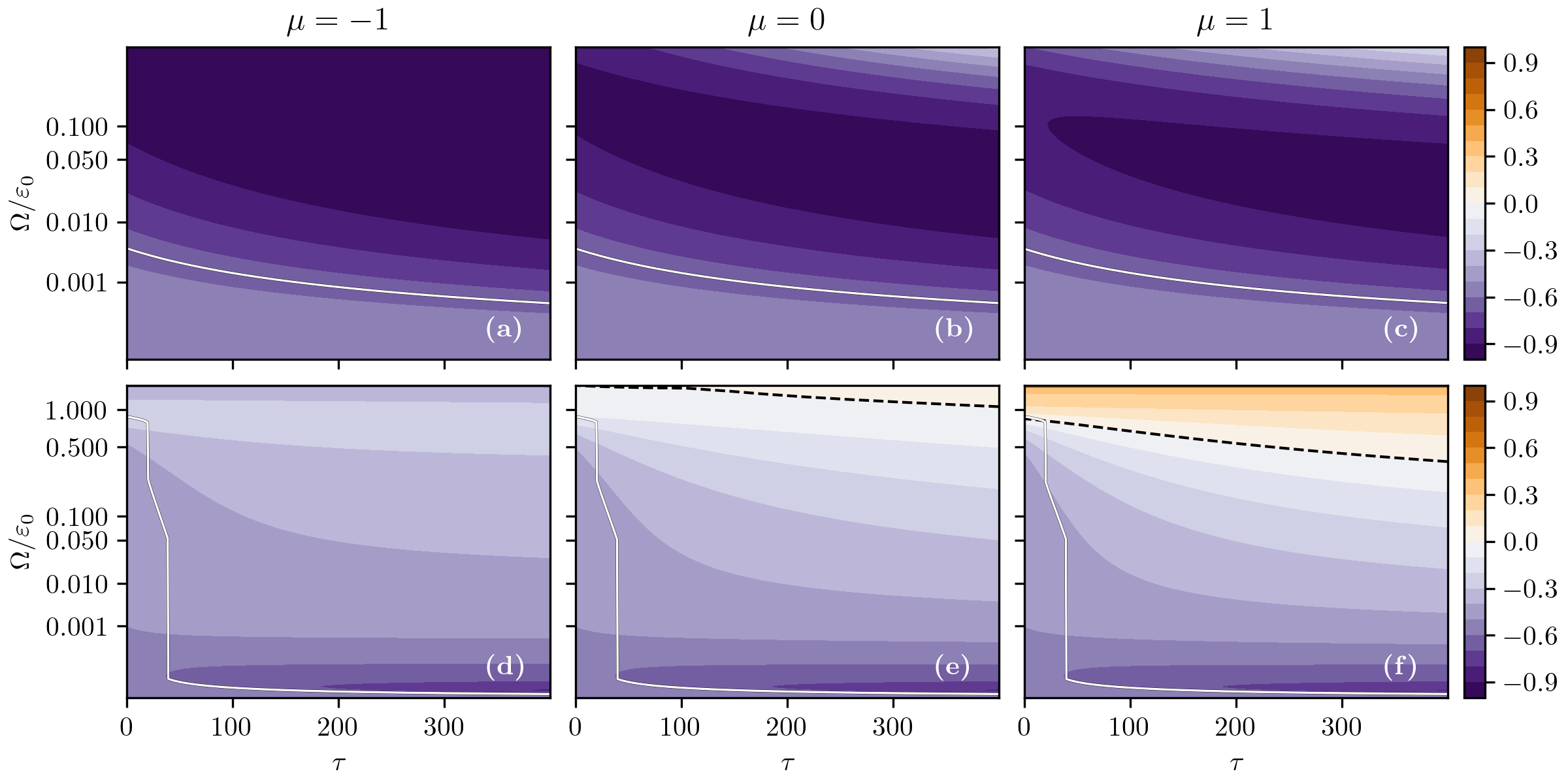}}
  \caption{Stokes parameter $Q$ of the synchrotron radiation as a function of dimensionless time $\tau$ and frequency $\Omega$ for the three cases of the initial spin orientation.
  The initial beam energy is $\varepsilon_0=10^{-2}$ (a--c) and $\varepsilon_0=10^4$ (d--f).}
  \label{fig:xi}
\end{figure}

Figure~\ref{fig:xi} shows the dependence of the Stokes parameter $Q$ on time and radiation frequency.
The most noteworthy feature is the significant depolarization of high-frequency radiation with $\Omega \sim \ec_0$, and even a change in its sign for an initially unpolarized beam or a beam polarized parallel to the field.
The impact of this effect is negligible if the initial beam energy is low (Fig.~\ref{fig:xi}a-c), because the emission of low-frequency photons dominates and the intensity decreases exponentially with increase in $\Omega$.
In this case,  $Q<0$ and the radiation is considerably polarized perpendicular to the magnetic field, regardless of the initial spin state of the beam.
In contrast, if the initial beam energy is large, $\varepsilon_0 \gg 1$, then the emission spectrum considerably broadens and its maximum shifted towards larger photon energies  $\Omega \sim \varepsilon_0$, as shown in {Fig.~\ref{fig:xi}(d-f).}
At the same time, emission of hard photons couses significant energy loss, and already at $\tau \sim 40$ intensity maximum shifts back to the low frequency region.

\begin{figure}
  \includegraphics[width=1.0\columnwidth]{{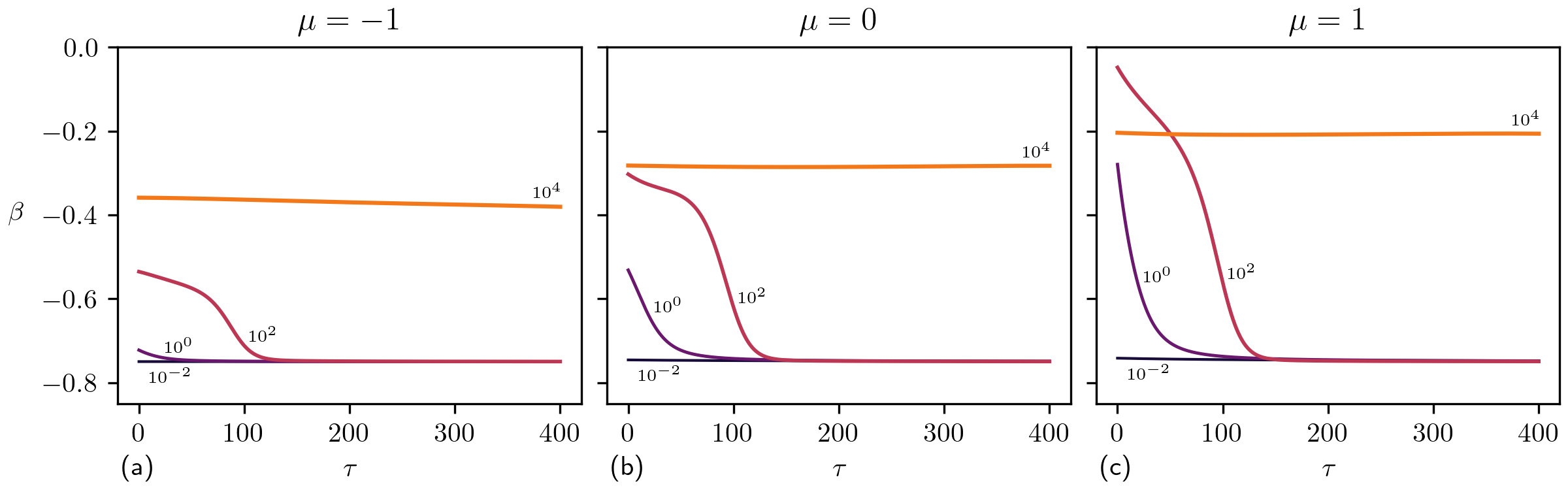}}
  \caption{Time evolution of the polarization parameter $Q$ defined via the total radiation intensity. The initial spin alignment is (a) antiparallel to the magnetic field ($\mu = -1$), (b) unpolarized, and (c) parallel to the field ($\mu = +1$). Curves correspond to different initial beam energies: $\ec_0=10^{-2}$; $10^{0}$; $10^{2}$; $10^{4}$.
  }
  \label{fig:xitot}
\end{figure}

When the spectrum maximum is located in the high energy range, the reduction in the magnitude of $Q$, and even its transition to positive values, leads to a significant decrease in the overall polarization of synchrotron radiation.
Figure~\ref{fig:xitot} shows the time dependence of the Stokes parameter $Q$, calculated using the intensity (\ref{intensity}) integrated over the radiation frequency $\Omega$ and substituted into the definition (\ref{stokes}).
In all cases the Stokes parameter is negative which corresponds to partial polarization in the direction perpendicular to the magnetic field.
Increase in the beam energy $\ec_0$ leads to a decrease in the magnitude of $Q$ while the beam energy remains large. 
This effect is most prominent if the electron spins are initially oriented parallel to the field, $\mu=+1$.
If $\ec_0 \gg 1$ holds, the process of self-polarization slows down due to the overall decrease of the synchrotron rates under this conditions.

\section{Conclusions}
In this work, we have conducted a theoretical analysis of radiative self-polarization of electrons in a hypothetical experimental setup involving an energetic electron beam propagating through a strong magnetic field.
Our focus was on the dependence of polarization phenomena on the dimensionless parameter $\ec$~(\ref{epsilon}), which is proportional to the product of the electron energy and the magnetic field strength.
To account for radiative energy losses and multiple synchrotron transitions, we numerically solved the balance equation for the energy distributions of the two spin components of the electron beam. 
This approach enabled us to investigate the evolution of both the spin alignment of the beam and the polarization of the emitted radiation.

As expected, the resulting degree of spin alignment increases with the mean beam energy $\ec_0$ in the regime $\ec_0 \ll 1$.
In contrast, for $\ec_0 \gg 1$, the dimensionless time required for self-polarization noticeably increases. 

At low energies, the beam radiation is primarily polarized perpendicular to the magnetic field, consistent with known results.
However, when $\ec_0 \gg 1$, the spectral maximum shifts towards the region $\Omega \sim \ec_0$, where the polarization of emitted photons substantially decreases.

The most notable case arises for electrons with spins aligned parallel to the magnetic field. 
In this regime, the Stokes parameter $Q$ may decrease in magnitude or even attain positive values, leading to significant or nearly complete depolarization of the synchrotron radiation.

\begin{acknowledgments}
To the memory of Dr.~Roman Kholodov, who passed away during the preparation of this work. 
\end{acknowledgments}


\end{document}